\documentclass[11pt]{article}
\input epsf.tex
\newdimen\SaveWidth \SaveWidth=\textwidth
\newdimen\SaveHeight \SaveHeight=\textheight
\advance\SaveWidth by -\textwidth
\advance\SaveHeight by -\textheight

\divide\SaveWidth by 2
\divide\SaveHeight by 2
\advance\hoffset by \SaveWidth
\advance\voffset by \SaveHeight

\def\ewviol{\rm {EW\!\!-\!breaking}}
\def\cpviol{CP ~\rm{violation}}
\def\cpviolng{CP ~\rm {violating}}
\def\susyviol{\slashword{SUSY}}

\let\badcite=\cite
\def\cite{~\badcite}

\def\slashchar#1{\setbox0=\hbox{$#1$}           
   \dimen0=\wd0                                 
   \setbox1=\hbox{/} \dimen1=\wd1               
   \ifdim\dimen0>\dimen1                        
      \rlap{\hbox to \dimen0{\hfil/\hfil}}      
      #1                                        
   \else                                        
      \rlap{\hbox to \dimen1{\hfil$#1$\hfil}}   
      /                                         
   \fi} 
    \def\slashword#1{\setbox0=\hbox{$#1$}        
  \dimen0=\wd0                                   
   \setbox1=\hbox{/} \dimen1=\wd1                
   \ifdim\dimen0>\dimen1                         
      \rlap{\hbox to \dimen0{\hfil\bf---\hfil}} %
      #1                                         %
   \else                                         
      \rlap{\hbox to \dimen1{\hfil$#1$\hfil}}    
      /                                          
    \fi}                                         %

\catcode`@=11
\newdimen\vbigd@men                             

\def\vbig#1#2{{\vbigd@men=#2\divide\vbigd@men by 2%
   \hbox{$\left#1\vbox to \vbigd@men{}\right.\n@space$}}}
\catcode`@=12

\catcode`@=11
\def\citenum#1{\csname b@#1\endcsname}
\catcode`@=12

\def\dofig#1#2{\centerline{\epsfxsize=#1\epsfbox{#2}}}
\begin{document}
\begin{titlepage}
\rightline{LBNL-45281}

\bigskip\bigskip

\begin{center}{\Large\bf\boldmath
Constraining  $\cpviolng$  phases of the MSSM\footnotemark \\}
\end{center}
\footnotetext{ This work was supported by the Director, 
Office of Science, Office
of Basic Energy Services, of the U.S. Department of Energy under
Contract DE-AC03-76SF0098.
}
\bigskip
\centerline{\bf I. Hinchliffe and N. Kersting}
\centerline{{\it Lawrence Berkeley National Laboratory, Berkeley, CA}}
\bigskip

\begin{abstract}

	Possible $\cpviol$ in supersymmetric (SUSY)
	extensions of the Standard Model (SM) is discussed. The consequences
	of $\cpviolng$ phases in the gaugino masses, trilinear soft \\
	supersymmetry-breaking terms and the $\mu$ parameter are explored.
	Utilizing the constraints on these parameters from electron and
	neutron electric dipole moments, possible $\cpviolng$ effects
	in $B$-physics are shown. A set of measurements 
	from the $B$-system which would overconstrain the 
	above $\cpviolng$ phases is illustrated.

\bigskip	

\end{abstract}

\newpage
\pagestyle{empty}

\end{titlepage}

\section{Introduction}
\label{sec:intro}

	The peculiar flavor structure of the Standard Model (SM) currently 
	has no generally-accepted explanation.
	 Included in this structure is the prediction that
	fundamental physics is not invariant under the operations of
	parity(P), charge(C), or their combination, CP.
	 More precisely, it is the
	distribution of complex numbers in the SM Lagrangian which
	lead to $\cpviol$ at an extremely small
\footnote{Jarlskog \cite{jarlskog}
has expressed this as follows: considering the interaction-basis 
up- and down-quark 
mass-matrices $m$, $m'$, one can form a rephase-invariant measure 
of $\cpviol$ equal
to $a \equiv 3 \sqrt{6} Det(C)/(Tr(C^2))^{3/2}$ where 
$iC \equiv [m m^\dagger, m' m'^\dagger]$. Then $-1 \le a \leq +1$ and 
$\cpviol \Leftrightarrow a \neq 0$. In the SM one finds 
$a \approx 10^{-7}$ }
but observable level \cite{jarlskog}. Currently experiments 
\cite{babar,belle,cdf,cleo2,d0,hera,lhc} are taking data or under
construction that will 
very precisely measure the Cabbibo-Maskawa-Kobayashi (CKM)
\cite{ckm} matrix elements 
believed responsible for $\cpviol$ in the SM. Any observed discrepancy with
the SM prediction indicates that the SM is incomplete in the flavor sector
and new physics must appear in the fundamental theory. 
An excellent candidate for new physics is Supersymmetry (SUSY) 
\cite{martin}. As SUSY introduces
 many more potentially large sources of $\cpviol$ in the mass matrices and
various field couplings, its inclusion requires that there be some
 suitable relationships among the parameters 
to reproduce the agreement of the existing levels of $\cpviol$ in 
$K$-decay and electric dipole moment (EDM) data with the SM predictions 
	\cite{susyphases}. The understanding of what $\cpviolng$ parameters
	are allowed in SUSY models therefore provides a constraint on such
	models.
	
	The simplest SUSY models can not satisfy the experimental 
	EDM bounds without either setting all SUSY $\cpviolng$ phases to
	zero or raising SUSY particle masses above $1~TeV$ 
\cite{polch,nath,kizu}. 
        Recent works have noted, however, that the supergravity-broken 
       MSSM with 
	$O(1)$ phases for the gaugino masses $M_i$, 
	triscalar coupling $A$, 
	and Higgs coupling parameter $\mu$, and no 
	flavor-mixing
	 beyond the standard CKM matrix can be consistent with existing limits
	 on the electric dipole moments (EDMs) of 
	the electron and neutron \cite{brhlik,ibra} . Given this, we may
	 now 
	investigate how the phases in this model will
	contribute to other $\cpviolng$ 
	observables and how new measurements can
	constrain the values of these phases. We choose to
	investigate several 
	$\cpviolng$ observables of the $B$-meson system, as $B$-processes
	occupy both a favorable theoretical and experimental position. On
	the theoretical side, uncertainties from non-perturbative QCD 
	are low due to the large mass of the $b$-quark 
	($\approx 4~GeV$)\cite{bmass}
	 relative to the energy scale
	 $\Lambda \approx 200~ MeV$\cite{hinch} characteristic of
	strong interactions and perturbation theory in 
	$\alpha_s$ is reliable. Furthermore, many SM $\cpviolng$ asymmetries
	in the $B$-system are small due to the suppression of CKM matrix
	elements involving the third generation; new $\cpviolng$ physics
	may be easily detectable. Experimentally, many dedicated facilities are already or
	will soon be
	generating large amounts of high-precision $B$-physics 
	data \cite{bphys}. From our analysis we will see that these
	experiments will be able to either precisely determine the above
	set of five phases or reject this particular model of
	$\cpviol$ altogether.

	We first briefly discuss the features of the 
	MSSM in Section \ref{sec:themodel} and review the 
	EDM constraints in Section \ref{sec:edm}. 
	In Section \ref{sec:bphys} we begin discussion of how the above phases
	enter observables in various sectors of $B$-meson physics through 
	$\cpviol$ in pure mixing, mixing and decay, and pure decay effects in the 
	processes $b \to s~\gamma$, $B_s^0 \to \phi~\phi, J/\psi~\phi$, and
	$B^- \to \phi~ K^-$ .
	Finally, we collect results and compare with the expected experimental
	sensitivities in Section \ref{sec:discuss}.
	Throughout this discussion we reserve most of the more 
	complicated formulae
	and analyses for the Appendix, to which we will refer the reader
	at the appropriate points.

\section{The Model}
\label{sec:themodel}

The setting for our calculations is the MSSM superpotential
\begin{equation}
\label{superpot}
{\bf W} =  {\bf y_u} \overline{u} Q H_2 +  {\bf y_d} \overline{d} Q H_1
  + {\bf y_e} \overline{e} L H_1 + \mu H_1 H_2
\end{equation}
with the chiral matter superfields for quarks and leptons $\overline{u}$
 , $\overline{d}$ , $Q$ , $\overline{e}$ , $L$ , and Higgs $H_{1,2}$
 transforming under $SU(3)_C \times SU(2)_L \times U(1)_Y$ as: 
\begin{equation}
 \begin{array}{lcccccccr}
	\overline{u}& \equiv& (\overline{3},1,-{2\over 3} ) &
           \overline{d}& \equiv& (\overline{3},1,{1\over 3}) &
                  Q&      \equiv& (3,2,{1\over 6}) \\
     
	\overline{e}& \equiv& (1,1,1)&   L& \equiv& (1,2,-{1\over 2})& \\
	H_1 & \equiv& (1,2,-{1\over 2})&     H_2 & \equiv& (1,2,{1\over 2})& \\

 \end{array}
\end{equation}

The $3 \times 3$ Yukawa matrices ${\bf y_{u,d,e}}$ couple the three
 generations of quarks and leptons, and the $\mu$ parameter couples the
two Higgs. In this discussion all
 flavor and gauge indices are implicit. The superpotential (\ref{superpot}) and general considerations of gauge invariance give the minimally supersymmetric $SU(3)_C \times SU(2)_L \times U(1)_Y$ Lagrangian 
\begin{equation}
\label{susylag}
{\cal L}_{SUSY} = {\cal L}_{kinetic} - \sum_i {dW\over d\phi^i}{dW^*\over d\phi^i} - { 1 \over 2 }\sum_a g^2_a (\phi^* T^a \phi)^2 
\end{equation}
where the kinetic terms have the canonical forms
\begin{equation}
	{\cal L}_{kinetic} = -D^\mu \phi^{\dagger i} D_\mu \phi_{i}
		- i \psi^{\dagger i} \overline{\sigma}^\mu D_\mu \psi_i
	 -{1 \over 4} F^{\mu \nu a} F_{\mu \nu a}
\end{equation}
with the indices $i$ and $a$  running over the scalar fields and gauge group
 representations, respectively. To this Lagrangian we add a  
 soft supersymmetry-breaking ($\susyviol$) piece 
\begin{equation}
\label{softsusy}
 \begin{array}{rrr}
{\cal L}_{\susyviol} =& (M_3 {\tilde g} {\tilde g } +
  				M_2 {\tilde W} {\tilde W} + 
	M_1 {\tilde B} {\tilde B} ) - ( {\bf a_u} \overline{u}
	 Q H_2 +  {\bf a_d} \overline{d} Q H_1
  + {\bf a_e} \overline{e} L H_1 )_{scalar} \\ 
    &  - (\sum_{X_i=Q,\overline{u},\overline{d},\overline{e}} 
   {\bf m^2_i} X_i X^{\dagger}_i - \sum_{i=1,2} m^2_{H_i} H_i^* H_i
 - b \mu H_1 H_2)_{scalar} + h.c. & \\
\end{array}
\end{equation}
where (${\tilde g}$, ${\tilde W}$, ${\tilde B}$) are the fermionic 
partners of the gauge bosons of  $SU(3)_C,SU(2)_L,U(1)_Y$, respectively,
 and  where {\it scalar} implies that only the scalar component of each 
superfield is used. SUSY-breaking of this type is characteristic 
of supergravity\cite{supergrav1,supergrav2}.

An additional simplification is to take the matrices ${\bf m^2_i}$ and 
${\bf a_{u,d,e} }$  in (\ref{softsusy}) along with the Yukawa matrices 
${\bf y_{u,d,e}}$ in (\ref{superpot}) to be simultaneously diagonal in the mass 
basis after electroweak symmetry breaking. This implies
\begin{equation} 
\label{diagonal}
 \begin{array}{lccr}
	{\bf m^2_i}& \equiv & m^2_0 {\bf 1} & \\  
        {\bf a_i }&   \equiv &  A {\bf y_i} & (i=u,d,e) \\ 
\end{array}
\end{equation}
Thus the only mixing between families is the usual CKM matrix since quark 
and squark mass matrices are diagonalized by the same rotation. 

The full Lagrangian ${\cal L}_{SUSY} + {\cal L}_{\susyviol}$ has six 
arbitrary phases imbedded in the parameters  $M_i (i=1,2,3)$, 
$\mu$ , $b$,  and $A$.  The phase of $b$ can immediately be defined to be zero by 
appropriate field redefinitions of $H_{1,2}$. A mechanism to set another
phase to zero utilizes the $U(1)_R$-symmetry of the Lagrangian which is broken by 
the supersymmetry-breaking terms, in particular the gaugino masses in (\ref{softsusy}).
We may perform a $U(1)_R$ rotation on the gaugino fields to remove
 one of the phases $M_i$.   For consistency 
with \cite{brhlik} we choose $M_2$ real ($ \phi_2 \equiv 0 $). Note that this  $U(1)_R$ 
transformation affects neither the phase of $A$, since having the Yukawa matrices
 ${\bf y_i}$ in (\ref{superpot}) be real fixes the phases of the same fields 
that couple to $A$, nor the phase of $\mu$, as
 having chosen
$\phi_b \equiv 0$ fixes the phases of $H_{1,2}$. Therefore the final set of physical phases
 we study is $\{ \phi_{1,3}, \phi_A, \phi_{\mu} \}$ .

\section{Electric Dipole Moment Constraints}
\label{sec:edm}

\begin{figure}[t]
\dofig{5.00in}{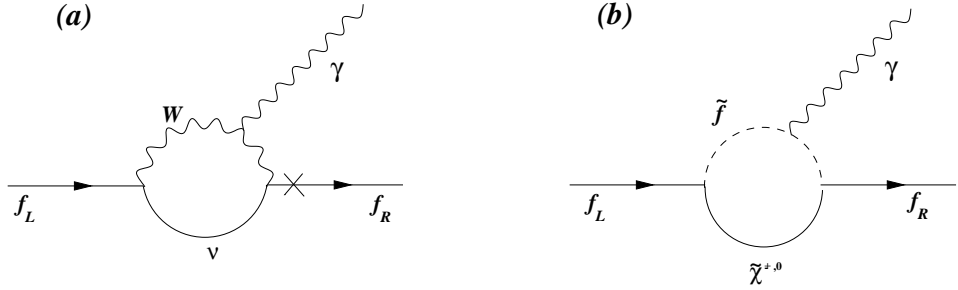}
\caption{1-Loop Contributions to the EDM in 
   {\bf \it (a)} the SM; the graph vanishes since the complex phases at
	the vertices cancel  {\bf \it (b)} in the MSSM, where 
	the requisite helicity-flip may be placed on either the chargino ($\tilde{\chi^{\pm}}$), 
        neutralino ($\tilde{\chi^0}$), or
	scalar ($\tilde{f}$) propagators to introduce $O(1)$ phases in the amplitude.
    }
\label{edmfig}
\end{figure}

The electric dipole moment (EDM) of an elementary fermion, a manifestly
$\cpviolng$ quantity, is the coefficient $d_f$ of the effective operator
$$
O_{edm} = - (i/2)\overline{f}  \gamma_5 \sigma_{\mu \nu} f F^{\mu \nu}
$$
where $F^{\mu \nu}$ is the electromagnetic field-strength tensor. 
The SM prediction of this coefficient vanishes at one loop
 (see Figure~\ref{edmfig}a) since the CKM phases from the two vertices cancel 
eachother. For the electron, $d_e$ even vanishes at two loops and the 
three-loop prediction is miniscule, of order $10^{-50}ecm$ \cite{donoghue}. For the 
neutron EDM, gluon interactions can give rise to a two-loop contribution to 
 $d_n$, but the result is still tiny at $d_n \leq 10^{-33}ecm$ 
\cite{shabalin}. The above predictions no longer hold if the QCD Lagrangian 
contains the $\cpviolng$ `$\theta$-term' $\theta \frac{g^2_{QCD}}{32 \pi^2}
 G^{\mu \nu} \tilde G_{\mu \nu}$, a potentially significant source of
$\cpviol$, but then the theory is consistent with the EDM limits only if 
$\theta<10^{-9}$\cite{crewther}; such a
fine-tuning is unnatural and henceforth we assume that $\theta=0$.
   
Although the amplitude of a typical supersymmetric contribution to the dipole 
moment (see Figure~\ref{edmfig}b) suffers relative to the SM contribution by a 
reduction factor of at least 
$(m_W/ \tilde{m})^2$ from SUSY particles of mass $\tilde{m}$ propagating in the 
loop, the imaginary piece of the SUSY amplitude could very well dominate over
that from the SM.
As previous calculations have shown \cite{polch,nath,kizu}, the diagrams in
 Figure~\ref{edmfig}b lead to EDMs in violation of the current limits of 
$d_e < 4.3 \times 10^{-27}ecm$ \cite{commins} and 
$d_n < 6.3 \times  10^{-25} ecm$ \cite{harris} unless SUSY particles are
heavier than $O(1) TeV$ or the $\cpviolng$ phases are less than $O(10^{-2})$. 
However, small phases in fact are not inevitable, for recently it has been 
pointed out \cite{brhlik,ibra} that a cancellation
can occur among the various SUSY diagrams, allowing $O(1)$ $\cpviolng$ 
phases to
be consistent with the current EDM experimental limits.

\begin{figure}[t]
\dofig{8.00in}{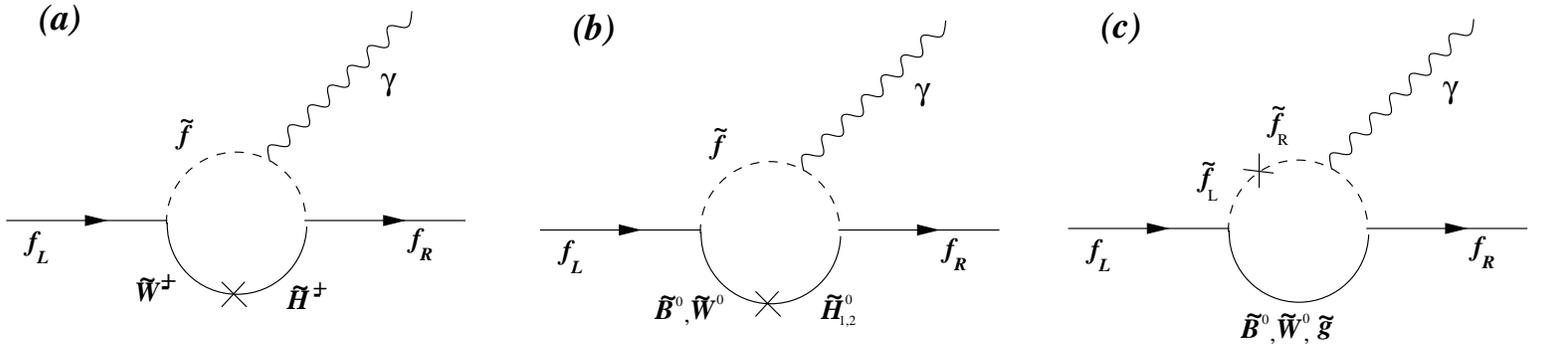}
\caption{Leading SUSY Contributions to the EDM: 
   {\bf \it (a)} charged wino ($\tilde{W}^{\pm}$) and  
   higgsino($\tilde{H}^{\pm}$) mixing provide the largest part of $d_f$ 
   which {\bf \it (b)} neutral wino($\tilde{W}^0$), 
   bino($\tilde{B}$), and higgsino($\tilde{g}^0$) mixing 
   partially cancels. {\bf \it (c)} exchange of a $\tilde{W}^0$, $\tilde{B}$,
   or gluino $\tilde{g}^0$ with mixing between the 
   scalar superpartners  ($\tilde{f}_{L,R}$) of the corresponding
   fermions $f_{L,R}$ could almost completely cancel against the other two
   diagrams. Note that the above processes actually occur via the 
   {\it mass} eigenstates $\tilde{\chi}^{\pm}$ and 
	$\tilde{\chi}^0$.
    }
\label{susyedm}
\end{figure}

The set of phases $\{ \phi_{1,3}, \phi_A, \phi_{\mu} \}$ enter 
in any diagram which involves mixing between the following
fields:
\begin{itemize}
\item Charginos ($\tilde{\chi}^{\pm}$): $\phi_\mu$ lies in the matrix 
	${\bf M_{\chi^{\pm}}}$ which mixes the set 
        $(\tilde{W}^{+},\tilde{H}^{+}_2,\tilde{W}^{-},\tilde{H}^{-}_1)$
\item Neutralinos ($\tilde{\chi}^{0}$): both $\phi_\mu$ and $\phi_1$ appear
	in the matrix ${\bf M_{\chi^{0}}}$ which mixes \\
	the set
        $(\tilde{W}^{0},\tilde{B},\tilde{H}^{0}_1,\tilde{H}^{0}_2)$
\item Scalars: terms in the Lagrangian such as 
$\mu^* {\bf y_u} {\overline{\tilde{u}}}{\tilde{u}} H^{0*}_1$ arising from 
the second term of (\ref{susylag}), and $\susyviol$-terms in 
(\ref{softsusy}) introduce $\phi_\mu$ and $\phi_A$, respectively, into 
the scalar mass insertions. The effect is particularly significant in 
$\tilde{t}$-mixing where the Yukawa matrices are large.
\end{itemize}

The SUSY diagrams which appear in Figure~\ref{edmfig}(b) fall into three
classes, shown in Figure~\ref{susyedm}. The charged wino-higgsino mixing 
diagram Figure~\ref{susyedm}(a) provides the dominant contribution to
$d_f$, with the phase of $\mu$ entering the amplitude as
expected. The neutralino-mixing diagram Figure~\ref{susyedm}(b) is 
numerically smaller than its charged counterpart, but it is of 
opposite sign and has the same dependence on $\phi_\mu$; therefore 
~\ref{susyedm}(a) and ~\ref{susyedm}(b) partially cancel. The
final type of diagram shown in Figure~\ref{susyedm}(c) has a more 
complicated phase dependence\footnote{we briefly outline this
dependence in the Appendix} which, in certain regions of parameter space
that are not fined-tuned, leads to a destructive interference
with the other two diagrams consistent with current experimental bounds 
on {\it both}  $d_n$ and $d_e$ \cite{brhlik}.

If $O(1)$ phases are then permissable, it is important to know whether or not 
other experimental observables can overconstrain these phases.
 We next show that the $B$ system 
alone provides enough observables to strongly constrain these phases.

\section{$B$ Physics Constraints}
\label{sec:bphys}

There are many reasons to consider the $B$-system in particular for 
measurements of $\cpviol$ beyond the SM, a few of which are:

\begin{itemize}
\item  SM contributions to the relevant observables are down
	by factors of small CKM matrix elements, so generic
	non-SM $\cpviolng$ physics should give a very clear signal. 
\item  uncertainties arising from strong interactions are small
	using Heavy Quark Theory \cite{bigi}
\item large amounts of data will be available in the near future
\end{itemize}

The $\cpviol$ in question can arise in any of three ways: through
$B^0-\overline{B}^0$ mixing, $B$-decay, or through their combination 
\cite{babar, bigi}. 

\subsection{CP violation in $B^0-\overline{B}^0$ mixing}

\begin{figure}[t]
\dofig{5.00in}{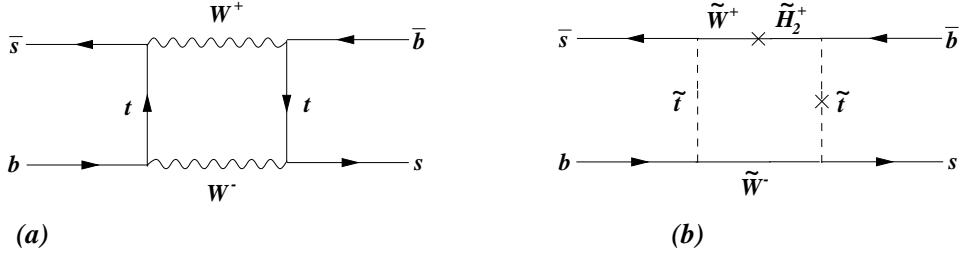}
\caption{Leading box diagrams in (a) the SM, and (b) the MSSM}
\label{susybox}
\end{figure}

At any given instant of time, the propagating meson states are 
 linear combinations 
 $|B_{L,H} \rangle \equiv p|B^0 \rangle \pm q|\overline{B}^0 \rangle $ which
evolve according to the  $2\times2$ Hamiltonian with dispersive and 
absorptive pieces ${\bf M}$ and ${\bf \Gamma}$, viz
\begin{equation}
\label{shrod}
        i \frac{d}{dt} \left( \begin{array}{c} B^0 \\ 
		\overline{B}^0  \end{array} \right)
	= \left( \begin{array}{cc} 
  M_{11} + \frac{i}{2} \Gamma_{11} & M_{12} + \frac{i}{2} \Gamma_{12} \\
  M_{12}^* + \frac{i}{2} \Gamma_{12}^* & M_{11} + \frac{i}{2} \Gamma_{11} \\
          \end{array} \right)
	\left(\begin{array}{c} B^0\\ \overline{B}^0  \end{array} \right)
\end{equation}
If we denote the time evolution of the states as
\begin{equation}
\label{eigenstates}
	|B_{L,H}(t) \rangle = 
          |B_{L,H}(0)e^{-i M_{L,H}}\rangle e^{-\frac{1}{2}\Gamma_{L,H}}
\end{equation}
then solving (\ref{shrod}) it follows\cite{babar} that
\begin{equation}
\label{cpcondition}
	\begin{array}{lcr}
 \cpviol \mathrm{~in~mixing} \Longleftrightarrow Im(\Delta M) \ne 0 & & 
         (\Delta M \equiv M_L-M_H) \\ 
	\end{array}
\end{equation}
Making use of the fact that 
$$ \begin{array}{lcr}
\Delta \Gamma \ll \Delta M &  & 
 (\Delta \Gamma \equiv \Gamma_L-\Gamma_H), \\
  \end{array}   
$$ 
which holds for both $B^0_d$ and $B^0_s$ \cite{babar,dunietz}, we
obtain the simplifications
\begin{equation}
\label{m12}
  \begin{array}{lcr}
	\Delta M \approx 2 |M_{12}| & & 
	\frac{q}{p}  \approx \frac{-|M_{12}|}{M_{12}} \\
  \end{array}
\end{equation}
The degree of $\cpviol$ in mixing is contained in the ratio $\frac{q}{p}$,
which from (\ref{m12}) is directly proportional to the phase factor in
the amplitude of the $\Delta S = 2$ box diagram (see Figure~\ref{susybox}).
In the SM, the box graph with internal top quarks in 
Figure~\ref{susybox}(a) dominates $Im(\Delta M)$; correspondingly, the strength
 of SM $\cpviolng$ mixing is CKM-suppressed by a factor
$Arg \left( V_{tb}V^*_{ts} \right)^2 \approx \eta \lambda^2 $ which
in the Wolfenstein approximation\cite{wolf} is $\approx 0.01$. 
Any  $\Delta S = 2$ contribution from new physics therefore has a generically
 large effect if it carries any phases with it at all. The
 leading supersymmetric chargino 
graph\footnote{We neglect gluino boxes in
the approximation that SUSY introduces negligable flavor mixing in the
 down-sector; boxes with additional Higgsinos are likewise suppressed by
 small Yukawa matrices. Further, we assume that the lightest stop dominates
the loops since it is usually the lightest squark.}
 in Figure~\ref{susybox}(b) provides the largest MSSM contribution to 
$Im(\Delta M)$. 
We have computed this box (see Appendix) and find it to dominate
significantly over the SM contribution. For example, at a typical point in SUSY parameter
 space, with $A \approx \mu$, $tan\beta=5$, and sparticle masses $\tilde m$ 
on the order of $O(2) \times M_W$, we obtain 
\begin{equation}
\label{q/p}
Im \left( q \over p \right) \approx 0.1 ~sin\phi_{\mu} cos\phi_A
\end{equation} 
which is an order of magnitude larger than the SM expectation and is directly
sensitive to  \\ 
$\cpviolng$ SUSY phases. However, it is difficult to directly
measure $Im \left( q \over p \right)$ through mixing effects alone; 
both time-dependent and time-integrated mixing effects are usually governed by
 $\Delta M$ which is still mostly real and SM-driven. We must therefore turn
 to other types of $\cpviol$ to constrain the MSSM phases.

\subsection{CP violation from mixing combined with decay}
 
\begin{figure}[t]
\dofig{5.00in}{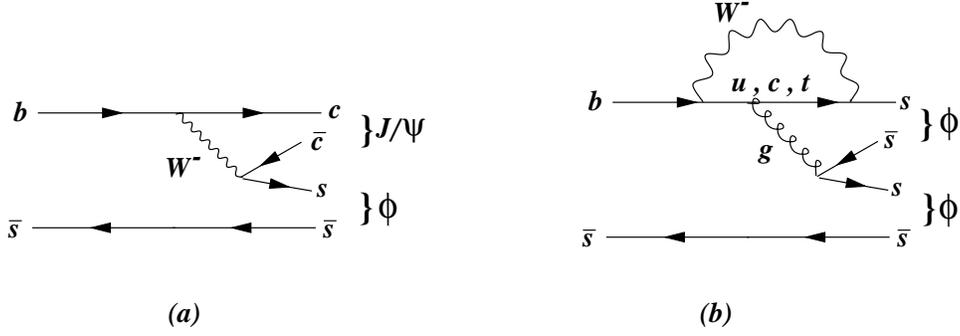}
\caption{Sizable $\cpviolng$ asymmetries arise from the processes 
(a) $\overline{B}^0_s \to J/\psi~\phi$, and
(b) $\overline{B}^0_s \to \phi\phi$ .
Only in the first case is hadronic uncertainty absent.}
\label{psiphi}
\end{figure}

When a particular final state $f_{CP}$ with definite CP quantum numbers is
accessible to the decays of a $B^0$ and $\overline{B}^0$ with amplitudes
$A_{f_{CP}}$ and $\overline{A}_{f_{CP}}$, respectively, the asymmetry
$$
 a_{f_{CP}} \equiv {{\Gamma (\overline{B^0} \to f_{CP}) -
			\Gamma (B^0 \to f_{CP})} \over
			 {\Gamma (\overline{B^0} \to f_{CP}) +
			\Gamma (B^0 \to f_{CP})}}
$$
in the limit of (\ref{m12}) becomes
\begin{equation}
\label{asymm}
	a_{f_{CP}} \approx {sin(\Delta M t)Im({q \over p} 
        \overline{\rho}_{f_{CP}}})
\end{equation}
where 
$$
\begin{array}{lr}
 \overline{\rho}_{f_{CP}} \equiv {\overline{A}_{f_{CP}} \over A_{f_{CP}}} &
 A_{f_{CP}}(B^0 \to f_{CP}) \equiv \sum_j A_j e^{i (\delta_j + \phi_j)} \\
\end{array}
$$
in general contains a dependence on both the weak phases $\phi_j$
and the strong interaction phases 
$\delta_j$ from each diagram contributing an amplitude $A_j$. 
In Figure~\ref{psiphi}, for example, the decay of a 
$\overline{B}_s^0$ to the final states $J/\psi~\phi$ and $\phi\phi$ may
either proceed directly ($\overline{B}_s^0 \to f_{CP}$) or by 
oscillating first 
($\overline{B}_s^0 \to  B_s^0 \to \overline{B}_s^0 \to f_{CP}$).
In the SM the asymmetries in both of these decays are tiny primarily due to the
 small $ \Delta S=2 $ mixing effects (see discussion following (\ref{m12}) 
 which  effectively give the upper bound
\begin{equation}
  a_{\phi \phi ~,~ J/\psi \phi} (SM) \le \eta \lambda^2 \approx 0.01
\end{equation}
 If SUSY exists then both of these asymmetries can be an order of magnitude
larger as is evident from
(\ref{q/p}) and (\ref{asymm}). Furthermore, the value of the quantity 
$\overline{\rho}_{f_{CP}}$ will differ from its value in the heavy quark
 expansion (HQE) by powers of $\Lambda_{QCD}/m_b$ only
\cite{chay, bigural,manohar} , so we set
the strong phases at $\delta_j = \pi$ with an uncertainty 
$(\Delta \delta)/\delta < 10\%$ \cite{desh,wshou}.  In the case of
 the decay to $\phi~\phi$
Figure~\ref{psiphi}(b), for example, $\overline{\rho}_{f_{CP}} \approx 1$,
whereas for the decay to  $J/\psi~\phi$ through the dominant tree-graph
Figure~\ref{psiphi}(a) carries no strong-phase dependence at all. 
From (\ref{q/p}) and (\ref{asymm}) it follows as a prediction of our model that 
\begin{equation}
 a_{J/\psi ~ \phi, \phi~\phi} \approx 0.1~ sin\phi_{\mu} cos\phi_A
\end{equation}
We now have two experimental $B$-physics signals of new $\cpviolng$ phases 
which constrain a combination of $\phi_{\mu}$ and $\phi_A$
independent from those which arise in the EDM bounds.

\subsection{CP violation in decay}
\label{subsec:cpdecay}

The charged $B$-mesons' decays can also serve to measure our set of phases.
The asymmetry in the decay to a final state $f$ is
\begin{eqnarray*}
  a_{f} \equiv {{1 - | \overline{A} /A | ^2}
         \over {1 + | \overline{A} /A | ^2}} \,\nonumber\\
	\,\nonumber\\
  A \equiv Amp(B^+ \to f) = \sum_j A_j e^{i (\delta_j + \phi_j)} \,\nonumber\\
  \overline{A} \equiv Amp(B^- \to \overline{f}) = 
             \sum_j A_j e^{i (\delta_j - \phi_j)}
\end{eqnarray*}
with $\delta_j$ and $\phi_j$ being the strong and weak phases, respectively for the diagram with modulus $A_j$ .
Rewritten in the form 
\begin{equation}
\label{bsg}
  a_{f} = {\sum_{i,j} A_i A_j sin(\phi_i - \phi_j) sin (\delta_i - \delta_j)
	\over \sum_{i,j} A_i A_j cos(\phi_i - \phi_j) 
	cos (\delta_i - \delta_j)}
\end{equation}
it is clear that a non-zero asymmetry requires that at least two diagrams 
contribute with different strong {\it and} weak phases.

\subsubsection{$b \to s ~\gamma$}
\label{subsub:bsg}

\begin{figure}[t]
\dofig{5.00in}{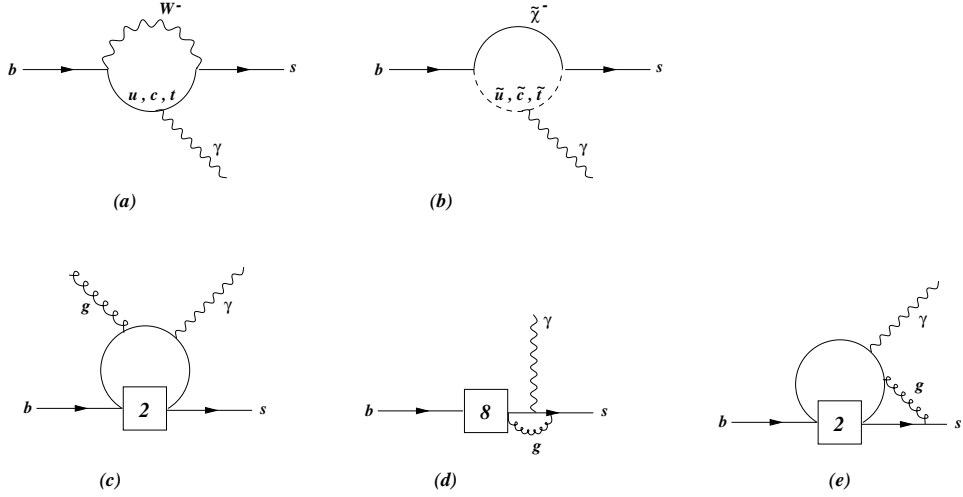}
\caption{Leading contributions to $b \to s~\gamma$ in 
(a) SM (b) MSSM. Graphs (c), (d), and (e) with the effective vertices 
$O_{2,8}$ play a significant role as well (mass insertions understood)}
\label{bsgfig}
\end{figure}

One of the interesting features of this mode is that any physical model 
that introduces new $\cpviolng$ phases can result in an asymmetry far greater
than that which the SM predicts; yet 
 it must not disturb the branching ratio(BR) for 
$b \to s \gamma$ which CLEO has measured\cite{CLEO}:
\begin{equation}
\label{branchbsg}
	BR(b \to s~\gamma) = (2.32~\pm~0.57~\pm0.35)\times 10^{-4}
\end{equation}
In Figure~\ref{bsgfig} we show the dominant diagrams; as in the case of the SUSY 
contributions to the EDMs studied above, the most important non-SM diagrams
for $b\to s\gamma$
involve the chargino loop in Figure~\ref{bsgfig}(b)\cite{bertolini}. Since the
 observed value of the branching ratio (\ref{branchbsg}) agrees very well with the
SM prediction , new decay channels are strongly constrained; accordingly, 
we follow the analysis of \cite{kagan} in carefully 
accounting for the higher-order graphs in  Figure~\ref{bsgfig}(c,d,e). 

The effective operators involved are
\begin{equation}
\begin{array}{l}
	O_2 \equiv \overline{s}_L \gamma_\mu q_L \overline{q}_L \gamma^\mu b_L
		\,\nonumber\\

	O_7 \equiv  {{e m_b}\over{4 \pi^2}} \overline{s}_L \sigma_{\mu \nu}
			F^{\mu \nu} b_R \, \nonumber \\
	O_8 \equiv {{g_s m_b}\over{4 \pi^2}} \overline{s}_L \sigma_{\mu \nu}
			G^{\mu \nu} b_R
\end{array}
\end{equation}
We leave the 
evaluation of these operators and all related calculations for the Appendix.

 In using these diagrams to compute observables it is important
 to take into account that the photon involved in the decay
 $b \to s\gamma$ is monochromatic but the photon in the observable
 $B \to X_s~\gamma$ has a variable energy. In addition to depending
 on the final state $X_s$, the photon energy is also a function of
 how the b-quark is bound inside the $B$-meson;
if the recoil energy of the b-quark is small, non-perturbative
effects arise for which no reliable models currently exist. These
considerations lead us to perform all calculations using a variable
outgoing photon energy, $E_\gamma$, which is bounded from below:
 $E_\gamma > (1- \xi) E_{max}$ , where $\xi$ is between 0 and 1 and $E_{max}$ 
is a model-dependent quantity. The actual dependence  on $\xi$ and $E_{max}$
in the computed asymmetry $a_{b \to s~\gamma}$ turns out to be negligable,
 as we demonstrate in the Appendix. Here we present the result    
$$
a_{b \to s \gamma} \approx 0.01~sin(\phi_\mu) 
$$
where as in the case $B^0_s \to \phi \phi$ the strong
 phase dependence is included in the coefficient and imparts a $10\%$ 
uncertainty to this prediction. The magnitude of the
asymmetry is admitedly not very large, however it is at least twice the SM 
prediction \cite{soares}.

\subsubsection{$B^- \to K^- \phi$}

\begin{figure}[t]
\dofig{5.00in}{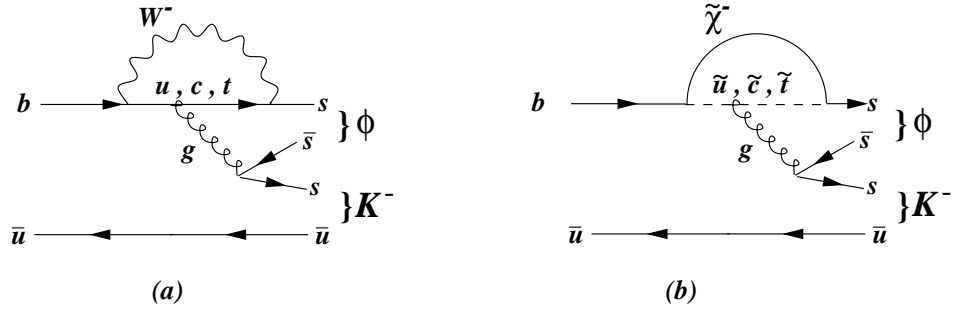}
\caption{$B^-_s \to K^- \phi$ in 
(a) the SM, where the $\cpviolng$ asymmetry is dominated by the CKM-suppressed 
 $u$-quark diagram and (b) the MSSM,
where the asymmetry is dominated by the relatively CKM-unsuppressed 
$\tilde{c}$ and $\tilde{t}$ graphs. Here mass insertions are understood.}
\label{bkphi}
\end{figure}

One particularly striking signature of the presence of the SUSY phases is in 
the decay \nolinebreak $B^- \to K^- \phi$. Here the flavor structure 
$b \to s~\overline{s}~s$ forbids
a SM tree graph, so the
leading SM graph is a penguin (defined `P'), as shown in
 Figure~\ref{bkphi}(a). 
The 
leading SUSY contribution in Figure~\ref{bkphi}(b) is also a
 type of penguin (hereafter designated a superpenguin, `SP'),
suppressed however by a factor $(m_W/{\tilde m})^2$ relative to the SM 
due to a squark propagating in the loop instead of a $W$-boson. 
So far the situation parallels the decay  $B^0 \to \phi~ \phi$ above,
but the major difference is that here the $\cpviol$ is necessarily 
{\it direct}. Referring back to (\ref{bsg}), with $\{i,j\}$ running
over $\{u,c,t,\tilde{u},\tilde{c},\tilde{t}\}$, we see that the asymmetry
receives contributions from 36 interference terms. However the 
imaginary parts of SP-SP interference terms
 are zero since the squarks in the loops of SP graphs are heavier than
the $b$-quark and do not give rise to absorptive phases in the
amplitude({\it i.e.} the $\delta$ 's are
zero in (\ref{bsg})). This
leaves SP-P and P-P terms. The latter, being purely SM terms, must
always involve at least one $u$-quark to get a nonzero weak phase 
($V_{tb}V^*_{ts} = -V_{cb}V^*_{cs} = -A \lambda^2$ is real) , whereas
the former need not involve $u$-quarks since the
weak phase difference necessary for a non-zero asymmetry can come from 
a SUSY coupling in $\tilde{c}$ or $\tilde{t}$ graphs.
Therefore the dominant SP-P terms will (in the notation of \ref{bsg}) have 
 $i \in \{\tilde{c},\tilde{t}\}$ and $j \in \{c,t\}$, leading over the P-P
terms by a factor $(V_{tb}V^*_{ts})/Im(V_{bu}V^*_{us}) \approx 
1/{\eta \lambda^2} \approx 100$. Therefore the P-P terms are 
negligable and the asymmetry is fundamentally due to the SUSY-SM 
interference. Assuming as before that the lightest $\tilde{t}$ dominates the
SP, the numerator of (\ref{bsg}) only contains the term due to
$\tilde{t}$-$c$ interference:
\begin{equation}
a_{B^- \to K^- \phi}   \approx 
  \frac{A_{c} A_{\tilde{t}} sin(\phi_c - \phi_{\tilde{t}})sin(\delta_c)}
 {A_{c}^2} 
\end{equation}
We now employ the one-loop perturbative calculation of the strong phase
$\delta_c$ from the $c$-quark loop \cite{wshou, desh}, and since 
$\phi_c \approx \eta \lambda^2$ is
 negligable and 
$\phi_{\tilde{t}}$ is essentially the same phase we calculated for
$b \to s~\gamma$\footnote{although the helicity structure of the
 ingoing $b$-quark and outgoing $s$-quark is L-R for $b \to s~\gamma$,
the L-L and R-R helicity structures of $B^-_s \to K^- ~\phi$ give
negligable contributions to the {\it phase} of the diagram} we obtain 
\begin{equation}
 a_{B^- \to K^- \phi} \approx   \left( {M_W \over \tilde m} \right)^2  
		sin(\phi_\mu) 
\end{equation}
Here we see that the absence of neutral flavor changing currents,
 the hierarchy of
the CKM matrix, and the pattern of CP quantum numbers all conspire in 
this case to give an asymmetry which is essentially zero in the SM yet for
 SUSY with typical sparticle masses can be as large as 30\%! 

\section{Discussion}
\label{sec:discuss}

We have seen that the $B$-system observables above provide many 
 constraints on the phases $\{\phi_A, \phi_{\mu} \}$
 in the model. We may classify the experiments by the size of the event
samples they are expected to provide: 
\begin{description}
\item[High Luminosity(H$\cal L$):]
	For example experiments at the LHC p-p collider, 
	producing a sample of order $10^{10}$ $B^0-\overline{B}^0$ pairs 
	per year \cite{lhc}
\item[Low Luminosity(L$\cal L$):]
 	 Experiments run at $e^+/e^-$ machines such as CESR(CLEO III), \\ 
		KEKB(Belle), and PEP-II(BaBar) as well as at hadronic 
	machines such as
	 Fermilab(CDF, D{\O}) and DESY(HERA-B: actually $e^+ p$) produce similar
	 samples of $B^0-\overline{B}^0$ pairs, \\ of order $10^7$ per year 
	  \cite{babar,belle,cdf,cleo2,d0}
\end{description}
   Any experiment which detects $N$ $B^0-\overline{B}^0$ pairs can resolve at
the 1~$\sigma$-level an  asymmetry $a_p$ for a process $p$ if
$$
	a_p >  \frac{1}{\sqrt{BR(p) N}}
$$
Using this criterion, the results in Table~\ref{results} summarizes
 the various modes
 and their asymmetries, comparing them to the experimental precision expected
on each asymmetry. The reader should note that the columns designated 
`H$\cal L$' and `L$\cal L$' refer to the {\it optimal} choice out of the
corresponding set of experiments. In some cases experiments in the same
set may have drastically different capabilities: for example the 
production of $B_s^0$ in $e^+e^-$ annihilation requires running on
 the $\Upsilon(5s)$ resonance, not currently possible 
 at BaBar or Belle, which run at the $\Upsilon(4s)$. Correspondingly the 
only observable in  Table~\ref{results} available to
these latter is the inclusive decay $b \to s~\gamma$ in the decay
of $B_d$-mesons. The decays of the daughter
mesons necessary for detection of events is taken into account in the
`$BR(p)$' column; for example, 
$BR(J/\psi \to (e^+~e^-,\mu^+~\mu^-)) \approx 12\%$ and
$BR(\phi \to K^+~K^-) \approx 50\%$. 
 More details on the individual capabilities of
each experiment may be found in \cite{bphys}.

\begin{table}
\caption{Predicted Asymmetries in the $B$-system and Experimental Error.
	$a_p$ is listed for both {\it high luminosity}(H$\cal L$) and 
{\it low luminosity}(L$\cal L$) experiments (see text for explanation)
in one year of running.
The strong phase uncertainty factors $\delta_{i} \approx 1\pm0.1$. For the
 explicit form of the function $f(\phi_\mu,\phi_A)$
 see the Appendix}
\label{results}
\begin{center}
\begin{tabular}{|l|l|l|l|l|}
\hline
$Process(p)$ & $BR(p)$ & $a_p$(predicted)  & $a_p$ (H$\cal L$)
	 & $a_p$ (L$\cal L$) \\
\hline
$b \to s~\gamma ~~~(B^0 \to X_s ~\gamma)$ & $2 \times 10^{-4}$ & 
	$0.01~sin(\phi_\mu)\delta_1$ &   $0.001$ & 0.02 \\
$B^0_s \to \phi~\phi$ &  $4 \times 10^{-5} \times (0.50)^2$ &
	$0.10~ f(\phi_\mu,\phi_A)\delta_2$ & $0.003$ & 0.09 \\ 
$B^0_s \to J/\psi~\phi$ & $10^{-3} \times (0.12)\times(0.50)$ &
	$0.10~ f(\phi_\mu,\phi_A)\delta_3$ &  $0.001$ & 0.03 \\
$B^-_s \to K^-~\phi$ & $10^{-5} \times (0.50)$ & $0.3~sin(\phi_\mu)\delta_4$
 &  $0.004$ & 0.12 \\
 \hline
\end{tabular}
\end{center}
\end{table}

Although HQE usually yields results perturbatively convergent in 
powers of $\frac{\Lambda}{m_b}$
, we allow for hadronic uncertainties at the level of $10\%$ on all
 observables in
 Table~\ref{results}  \cite{kagan,bigi,wshou}. The parameters describing this are the $\delta_i~(i=1..4)$ which are in general completely independent for the
observables in question. A valid test of the model requires a $10\%$
measurement of the various asymmetries. Disagreement at higher
precision could be ascribed to uncertainties in the strong dynamics.

From the table, we see that L$\cal L$-experiments can only 
contribute in the asymmetries in \\ $B^-_s \to K^-\phi$ and  
$B^0_s \to J/\psi~\phi$. Combined with the H$\cal L$ measurements of all
of the decays studied, a determination of $\phi_\mu$ and  $\phi_A$
will be possible. Combined with the linear combination of phases which electron
 and neutron EDM's constrain  (see Appendix) this provides
a complete determination of the set of the phases 
$\{ \phi_{1,3}, \phi_A, \phi_{\mu} \}$ studied in this
model of $\cpviol$.

In summary, the possibility that the phase structure of the MSSM 
extends beyond the
trivial one where $\cpviol$ is confined to the CKM matrix leads not
only to the requirement that SUSY phases respect the present EDM bounds,
but also that the range of phases consistent with these bounds agrees
with the values which can be extracted from the various $B$-system 
asymmetries considered above. Collectively , measurements of these 
asymmetries at present and future $B$-physics experiments 
will either determine the phases or rule out this 
particular SUSY model.  

\section*{Acknowledgments}
 This work was supported by the Director, Office of Science, Office
of Basic Energy Services, of the U.S. Department of Energy under
Contract DE-AC03-76SF0098.

\section*{Appendix}

\subsection*{1. Cancellation of the EDMs}
	That the contributions to the EDM in Figure~\ref{susyedm}(a,b) tend
	to cancel in a way dependent only on $sin(\phi_\mu)$ is evident from
	the form of the chargino and neutralino mixing matrices:
$$
{\cal L} ~\supset~ \tilde{\chi}^{\pm~T}{\bf M_{\tilde{\chi}^\pm}}
	 \tilde{\chi}^{\pm}
		+ \tilde{\chi}^{0~T}{\bf M_{\tilde{\chi}^0}}
		\tilde{\chi}^{0}
$$
	where
$$
{\bf M_{\tilde{\chi}^\pm}} \approx \left( 
     \begin{array}{cccc}
      0 & 0 & M_2 & 0 \\  0 & 0 & 0 & \mu \\  M_2 & 0 & 0 & 0 \\
      0 & \mu & 0 & 0 \\
     \end{array} \right)
~~~{\bf M_{\tilde{\chi}^0}} \approx \left( 
     \begin{array}{cccc}
      M_1 & 0 & 0 & 0 \\  0 & M_2 & 0 & 0 \\  0 & 0 & 0 & -\mu \\
      0 & 0 & -\mu & 0 \\
     \end{array} \right)
$$
assuming $ M_W \ll M_2,~\mu $. 

The graph in Figure~\ref{susyedm}(c) (where $f$ is, say, an electron)
receives phases from three sources:
\begin{itemize}
\item the $U(1)$ propagator carries a factor $e^{i \phi_1}$.

\item the scalar mass insertion has a SUSY piece given from the first term
 of (\ref{susylag}) which has the form 
${\bf y_e} \mu^* \tilde{\overline{e}}\tilde{e} H_2^{0*}$ which after
$\ewviol$ becomes ${\bf y_e} \mu^* \tilde{\overline{e}}\tilde{e}~v~sin\beta$
\item the scalar mass insertion also has a $\susyviol$ piece 
(see (\ref{softsusy}) ) of the form
$ A {\bf y_e} \tilde{\overline{e}}\tilde{e} H_1^{0}$ which becomes
 $A {\bf y_e}\tilde{\overline{e}}\tilde{e}~v~cos\beta$ after $\ewviol$.
\end{itemize}
Putting the above pieces together, the imaginary piece of the 
neutralino graph in Figure~\ref{susyedm}(c) carries a phase dependent factor
$(|A|sin(\phi_A + \phi_1) + |\mu|~ tan\beta sin(\phi_1 - \phi_\mu))$.
Likewise, the corresponding graph for the neutron in the $SU(6)$
model\footnote{for alteratives, see \cite{bartl}}  where 
$d_n = 1/3 (4 d_d - d_u)$ obtains a factor similar to the
electron case, with the replacement $\phi_1 \to \phi_3$; the numerical
demonstration that these diagrams can nearly cancel is given in \cite{brhlik}.

\subsection*{2. Calculating the $B_s^0-\overline{B}_s^0$ Box}

We follow the notation of \cite{bertolini} in calculating the contribution
to $M_{12}$ from the chargino box in Figure~\ref{susybox}(b):
$$
\Delta M_{\tilde{\chi}^\pm} = {\alpha_w^2 \over 16}
	\sum_{h,k=1}^6 \sum_{i,j=1}^2 {1 \over m^2_{\tilde\chi^\pm}}
	(G^{jkb}_{UL} - H^{jkb}_{UR})(G^{*iks}_{UL} - H^{*iks}_{UR})
(G^{ihb}_{UL} - H^{ihb}_{UR})(G^{*jhs}_{UL} - H^{*jhs}_{UR})G'_{ijkh}
$$
where $G$ and $H$ are gauge and Higgs vertices and the
form factor $G'$ depends on the masses of the particles involved.
We make the assumption that the lightest chargino and
lightest stop dominate the loop, and that $M_W \ll M_2,~\mu$. The final
 result is
$$
Im(M_{12}) \approx \left({2 M_W \over \tilde{m}} \right)^3
	\left( 
  \frac{sin\phi_\mu (|A| sin\beta~cos\beta~cos\phi_A-
       |\mu|cos^2\beta~ y_t~ cos\phi_\mu)}
       {\sqrt{|A|^2sin^2\beta + |\mu|^2cos^2\beta - 
	|A||\mu|sin\beta~cos\beta~y_t cos(\phi_A + \phi_\mu)}} \right)
$$
which assumes a simpler form for typical points in parameter space,
as noted above (\ref{q/p}).
In keeping with prior notation, this defines
$$
 f(\phi_\mu,\phi_A) \equiv  \frac{sin\phi_\mu (|A| sin\beta~cos\beta~cos\phi_A-
       |\mu|cos^2\beta~ y_t~ cos\phi_\mu)}
       {\sqrt{|A|^2sin^2\beta + |\mu|^2cos^2\beta - 
	|A||\mu|sin\beta~cos\beta~y_t cos(\phi_A + \phi_\mu)}}
$$

\subsection*{3. The Operators $O_{2,7,8}$}
We again follow  \cite{bertolini} in calculating the coefficients $C_{7,8}$
 of the operators $O_{7,8}$ from chargino loops:
$$
\begin{array}{ll}
C_{7,8{\tilde{\chi}^\pm}} = & {{\alpha_w \sqrt{\alpha}} \over {2 \sqrt{\pi}}}
	\sum_{k=1}^6 \sum_{j=1}^2 {1 \over m^2_{\tilde{u}_k}}
	(G^{jkb}_{UL} - H^{jkb}_{UR})(G^{*jks}_{UL} - H^{*jks}_{UR})
	(F_{1,jk} + e_U F_{2,jk}) \\ 
	& - H^{jkb}_{UL}(G^{*jks}_{UL} 
	  -H^{*jks}_{UR}){m_{\tilde\chi^\pm_j} \over m_b}
        (F_{3,jk} + e_U F_{4,jk}) \\
\end{array}
$$
where $F_{1,2,3,4}$ are form factors and $C_7$ is obtained from the
 above by setting $\alpha = e^2/(4 \pi)$ and
$e_U = 2/3$; for $C_8$  $\alpha = g^2/(4 \pi)$ and $e_U = 0$.
Including the SM contributions given in \cite{kagan}, we obtain
$$
\begin{array}{lll}
	C_2 & \approx & 1.11 \\
	C_7 & \approx & -0.31-0.19~e^{i \phi_\mu} \\
	C_8 & \approx & -0.15-0.14~e^{i \phi_\mu} \\
\end{array}
$$
Note that $C_2$, the real part of $M_{12}$, is not significantly affected
by SUSY.

\subsection*{4. Soft Photons in $b \to s~\gamma$}

In \ref{subsub:bsg} we noted that the energy dependence of the outgoing
photon could have a significant effect on the BR and asymmetry. Here
we quote the expression in \cite{kagan} for the BR and asymmetry as
functions of $C_{2,7,8}$ and $\xi$ ($E_\gamma > (1-\xi)E_{max})$:
$$
\begin{array}{lll}
	{\mathrm BR} (B \to X_s ~\gamma) & \approx &
	2.57 \times 10^{-3} K_{NLO}(\xi)
	 \times {\mathrm BR} (B \to X_c ~e ~\nu)/10.5\% \\
{\mathrm where}  \\
 K_{NLO}(\xi) & \equiv & \sum_{i \leq j=2,7,8}
	\left(  k_{ij}(\xi)Re(C_i C_j^*) 
	+ k_{77}^{(1)}(\xi)Re(C_7^{(1)} C_7^*)  \right) \\

a_{b \to s\gamma} & = & \frac{1}{|C_7|^2} 
	\left( a_{27}(\xi) Im(C_2 C_7^*) + a_{87}(\xi)Im(C_8 C_7^*)
	+  a_{28}(\xi)Im(C_2 C_8^*) \right) \\
\end{array}
$$
 
\begin{table}
\caption{Dependence of the asymmetry in $b \to s~\gamma$ computed in the parton
model and Fermi-Motion model on the minimum energy of the soft photon.}
\label{xitable}
\begin{center}
\begin{tabular}{|c|c|c|c|}
\hline
$\xi$ &  $(1-\xi)E_{max}~~(GeV)$ & $|a_{parton}|$ & $|a_{fermi}|$ \\
\hline
1.00 &  0  &  0.008 & 0.008 \\
0.30 &  1.85 & 0.010 & 0.010 \\
0.15 & 2.24 & 0.011 & 0.012 \\
\hline
\end{tabular}
\end{center}
\end{table}

\begin{table}
\caption{Definitions of the $\xi$-Dependent Coefficients. `(p)' refers
to the `parton model' and `(f)' includes `Fermi motion'. }
\label{xiptable}
\begin{center}
\begin{tabular}{|c|c|c|c|c|c|c|}
\hline
$\xi$ &  $a_{27}^{(p)}$ & $a_{87}^{(p)}$  &  $a_{28}^{(p)}$ &
  $a_{27}^{(f)}$ &  $a_{87}^{(f)}$ &  $a_{28}^{(f)}$ \\
\hline
1.00 &  1.06  & -9.52 & 0.16 & 1.06 & -9.52 & 0.16 \\
0.30 &  1.17  & -9.52 & 0.12 & 1.23 & -9.52 & 0.10 \\
0.15 &  1.31  & -9.52 & 0.07 & 1.06 & -9.52 & 0.04 \\
\hline
\end{tabular}
\end{center}
\end{table}

The $\xi$-dependent quantities are listed in Table \ref{xiptable} which
we paraphrase from \cite{kagan}; using this, it is straightforward to 
explicitly calculate
the BR and asymmetry for the values of $C_{2,7,8}$ given above. The
dependence is insignificant for a wide range of $\xi$ and in two
different models (see Table~\ref{xitable}).

\end{document}